# Origin of Strong Magnetic Fields of Magnetars

Qiu-he Peng[1,2,*], Jie Zhang[2], Men-quan Liu[2], Chich-gang Chou[3]

[1]Department of Astronomy, Nanjing University, China
[2]Department of Physics, West-China Normal University, China
[3]National Astronomical Observatory, Chinese Academy of Sciences, China

**Abstract**  Since there is $^3P_2$ neutron superfluid in neutron star interior, it can be treated as a system of magnetic dipoles. Under the presence of background magnetic field, the magnetic dipoles tend to align in the same direction. When the temperature is lower than $10^7$K, the strong magnetic fields of the magnetars may originate from the induced magnetic moment of the $^3P_2$ neutron Cooper pairs in the anisotropic neutron superfluid. And this gives a convenient explanation of the strong magnetic field of magnetars.

## 1. Introduction

It is generally believed that there is very strong magnetic field, $B > 10^{12}$ gauss, for most of the neutron stars (e.g. Shapiro and Teukolski, 1983). There probably are magnetars with strong magnetic field strength exceeding the quantum critical threshold, $H_{cr} = 4.414 \times 10^{13}$ gauss (Duncan & Thompson 1992; Paczynski 1992; Usov 1992; Thompson & Duncan 1995,1996). Anomalous X-ray Pulsars (AXPs) and Soft Gamma Repeaters (SGRs) are classes of candidates to magnetars (e.g. Kouvelliotou et al. 1998, 1999; Hurley et al. 1999; Mereghetti & Stella 1995; Wilson etal. 1999; Kaspi et al. 1999). The magnetic field of the magnetars may be so strong as to reach two orders of magnitude above the quantum critical threshold $H_{cr}$.

It is believed that the strong magnetic field of neutron stars originates from the collapse of the core of a supernova with the conservation of magnetic flux (e.g. Ostriker & Gunn, 1969). However, it is rather difficult to get the strong magnetic field such as those expected for some of the pulsars with $B > 10^{13}$ gauss, especially for the strong field strength of magnetars by the collapse process alone.

Several models for the origin of magnetars have already been proposed. Ferrario & Wickrammasinghe(2005) suggest that the strong magnetic field of magnetars is descended from their stellar progenitor with high magnetic field core.

Iwazaki(2005) proposed that the huge magnetic field of magnetars is some color ferromagnetism of quark matter. Recently Vink & Kuiper (2006) suggest that magnetars may originate from the rapid rotating proto-neutron stars.

In this paper we propose a new idea for the origin of the magnetars. The strong magnetic fields of the magnetars may originate from the induced magnetic moment of the $^3P_2$ neutron Cooper pairs in the anisotropic neutron superfluid.

## 2. Anisotropic $^3P_2$ Neutron Superfluid

There are two relevant regimes of neutron superfluid in the neutron star interior: one is the isotropic $^1S_0$ neutron superfluid with the critical temperature is $T_\lambda$ ($^1S_0$)≈$1\times10^{10}$K in a density range $1\times10^{10} < \rho$ (g/cm$^3$) $<1.6\times10^{14}$. The other is the anisotropic $^3P_2$ neutron superfluid with a wide density range $1.3\times10^{14} < \rho$ (g/cm$^3$) $< 7.2 \times10^{14}$. The critical temperature $T_\lambda(^3P_2)$ is

$$T_\lambda(^3P_2) = \Delta_{\max}(^3P_2)/1.78k \approx 3.25\times10^8 K. \quad (1)$$

We note that the energy gap $\Delta(^3P_2)$ is almost a constant about the maximum with an error less than 3% in a rather wide density region $3.3\times10^{14} < \rho$ (g/cm$^3$) $< 5.2\times10^{14}$ ( see Fig. 8 of Elgagøy et al. 1996, but we neglect the F state of neutron Cooper pair here).

A $^3P_2$ neutronCooper pair has a spin angular momentum with a spin quantum number, $\sigma =1$. The magnetic moment of the $^3P_2$ neutronCooper pair is twice that of the abnormal magnetic moment of a neutron, $2\mu_n$ in magnitude, and its projection on the external magnetic field (z-direction) is $\sigma_z \times (2\mu_n)$, $\sigma_z =1,0,-1$, where

$$\mu_n = -0.966\times10^{-23} \text{ erg/gauss}$$

It is interesting to note that the behavior of the $^3P_2$ neutron superfluid is very similar to that of the liquid $^3$He at very low temperature (Leggett 1975):
1) The projection distribution for the magnetic moment of the $^3P_2$ neutronCooper pairs in the absence of external magnetic field is stochastic, or "Equal Spin Pair" (ESP) phase similar to the A- phase of the liquid $^3$He at very low temperature (Leggett 1975). The $^3$P$_2$neutron superfluid is basically isotropic without significant magnetic moment in the absence of the external magnetic field. We name it as the A- phase of the $^3$P$_2$neutron superfluid similar to the A phase of the liquid $^3$He at very low temperature (Leggett 1975).

2) However, the projection distribution for the magnetic moment of the $^3$P$_2$ neutron Cooper pairs in the presence of external magnetic field is not stochastic. The number of $^3$P$_2$ neutron Cooper pair with paramagnetic moment is more than the ones with diamagnetic moment. Therefore, the $^3$P$_2$neutron superfluid has a whole induced paramagnetic moment and its behavior is anisotropic in the presence of external magnetic field. We name it as the B- phase of the $^3$P$_2$ neutron superfluid similar to the B phase of the liquid $^3$He at very low temperature (Leggett 1975).

The full story of anisotropic $^3$P$_2$ neutron superfluid is more complicated when magnetic interaction is introduced. We consider this problem from the effective field theory point of view (Feng & Jin 2005). And for simplicity we only consider the zero order approximation. It is possible that the induced magnetic field of the $^3$P$_2$superfluid will boost the neutron star magnetic field to that of magnetars, whenever the stellar internal temperature is below the phase transition temperature. We will discuss this in see section V.

## 3. Induced Paramagnetic Moment of the $^3P_2$ Neutron Superfluid in the B-phase

The induced paramagnetic moment of the $^3$P$_2$neutron superfluid in the B-phase may be simply estimated as follows: the system of the $^3$P$_2$neutronCooper pairs with spin quantum number, s = 1, may be treated as a Bose-Einstein system. It obeys the Bose-Einstein statistics.

A magnetic dipole tends to align in the direction of the external magnetic field. The $^3$P$_2$ neutron Cooper pair has energy $\sigma_z \times 2\mu_n B$ ($\sigma_z$ =1,0,-1) in the applied magnetic field due to the abnormal magnetic moment of the neutrons. Here $\mu_n$ is the absolute value of the magnetic moment of a neutron, we use this convention from now on. B is the background magnetic field if there is no magnetic interaction. While in the effective field frame, B is the total magnetic field. Detailed discussion will be presented in section V. We denote the number density of $^3$P$_2$neutronCooper pairs with spin projection $\sigma_z$ =1,0,-1 by $n_1$, $n_0$ and $n_{-1}$ respectively. Their relative ratios are

$$\frac{n_{-1}}{n_0} = e^{2\mu_n B/kT}, \quad \frac{n_{+1}}{n_0} = e^{-2\mu_n B/kT}, \quad (2)$$

$$n_{-1} + n_0 + n_{+1} = n_n(^3P_2) \quad (3)$$

The difference of the number density of $^3$P$_2$ neutron Cooper pairs with paramagnetic and diamagnetic moment is

$$\Delta n_{\mp} = n_{-1} - n_{+1} = n_n f(\frac{\mu_n B}{kT}), \quad (4)$$

$$f(x) = \frac{2\sin h(2x)}{1 + 2\cos h(2x)}. \quad (5)$$

The Brillouin function, $f(\mu_n B/kT)$, is introduced to take into account the effect of thermal motion We note that $f$ It is an increasing function, in particular, $f(x) \approx 4x/3$, for $x \ll 1$ and $f(x) \to 1$, when $x \gg 1$. $f(\mu_n B/kT)$ increase with decreasing temperature. And this is the mathematical formula for the B-phase of the $^3$P$_2$ superfluid.

A relevant question is how many neutrons have been combined into the $^3$P$_2$ Cooper pairs? The total number of neutrons is given by $N = V \int_0^{k_F} \frac{d^3k}{(2\pi)^3}$ (Here $k$ is a wave vector, $p = \hbar k$, $p$ is the momentum).The neutrons combined into the $^3$P$_2$ Cooper pairs are just in a thin layer in the Fermi surface with thickness $k_\Delta$, $\hbar k_\Delta = \sqrt{2m_n \Delta_n(^3P_2)}$. Then we have $\delta N \approx \frac{V}{2\pi^2} k_F^3 \frac{k_\Delta}{k_F}$ ($k_\Delta \ll k_F$).

Thus, the fraction of the neutrons that combined into the $^3$P$_2$ Cooper pairs is

$$q = \frac{\delta N}{N} \approx 3\frac{k_\Delta}{k_F} \quad (6)$$

Here, we would like to emphasize that the energy gap, Δ, is the binding energy of the Cooper pair rather than a variation of the Fermi energy due to a disturbance or due to the variation of particle number density (In the latter case, it is easy to misunderstand to get $q \approx 3\frac{\delta k}{k_F} = \frac{3}{2}\frac{\delta E}{E_F} = \frac{3}{2}\frac{\Delta}{E_F}$ or $q = \frac{3}{8\pi^3}\frac{\Delta}{E_F}$). Therefore, we think the eq.(6) is a proper estimation.

For the non–relativistic neutron gas ($E_F = \hbar^2 k_F^2 / 2m_n$), the fraction of the neutrons that combined into the $^3$P$_2$ Cooper pairs is

$$q = \frac{4\pi p_F^2 [2m_n \Delta(^3P_2)]^{1/2}}{(4\pi/3)p_F^3} = 3[\frac{\Delta(^3P_2)}{E_F}]^{1/2} \quad (7)$$

The Fermi energy of the neutron system may be calculated by the formula

$$E_F = \frac{1}{2m_n}(\frac{3}{8\pi})^{2/3} h^2 N_A^{2/3}(Y_n\rho)^{2/3} \approx 60(\frac{\rho}{\rho_{nuc}})^{2/3} MeV \quad (8)$$

The energy gap of the anisotropic neutron superfluid is $\Delta(^3P_2) \sim 0.045$ MeV (Elgarøy et al. (1996)), q ~ 8.7%. Thus, the total number of the $^3$P$_2$ Cooper pairs is

$$N_n(^3P_2\_pair) \approx q N_A m(^3P_2)/2. \quad (9)$$

Therefore, the total difference of the $^3P_2$ neutron Cooper pair number with paramagnetic and diamagnetic moment is

$$\Delta N_\mp = N_n(^3P_2\_pair) f(\frac{\mu_n B}{kT}) = \frac{1}{2} N_A m(^3P_2) q f(\frac{\mu B}{kT}) \quad (10)$$

The total induced magnetic moment, of the anisotropic neutron superfluid is

$$\mu_{pair}^{(tot)}(^3P_2) = 2\mu_n \times \Delta N_\mp = \mu_n N_A m(^3P_2) q f(\mu_n B/kT). \quad (11)$$

Where $m(^3P_2)$ is the mass of the anisotropic neutron superfluid in the neutron star, $N_A$ is the A'vgadro constant. The magnetic moment with the dipolar magnetic field is $|\mu_{NS}| = B_p R_{NS}^3 / 2$ (Shapiro and Teukolski, 1984). Here $B_p$ is the polar magnetic field strength and $R_{NS}$ is the radius of the neutron star. The induced magnetic field is then

$$B^{(in)} = \frac{2\mu_{pair}^{(tot)}(^3P_2)}{R_{NS}^3} = \frac{2\mu_n N_A m(^3P_2)}{R_{NS}^3} q f(\mu_n B/kT). \quad (12)$$

Or

$$B^{\mathbf{Y in Y}} = B^{(in)}{}_{\max(^3P_2)} \cdot f\left(\frac{\mu_n B}{kT}\right) \quad (13)$$

The induced magnetic field for the anisotropic neutron superfluid increases with decreasing temperature (as shown in the Fig.1).

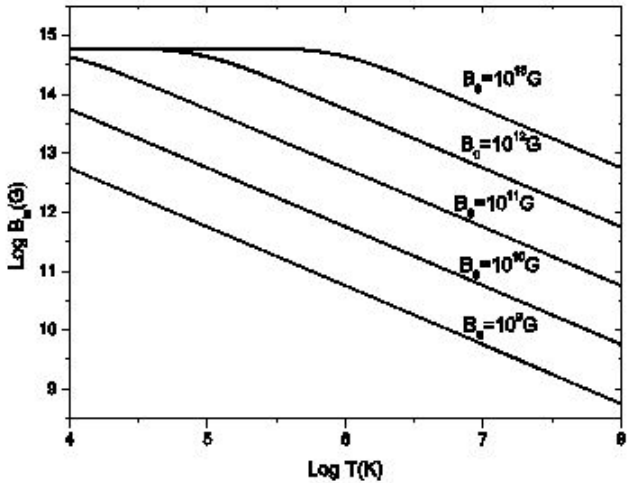

**Figure 1.** Variation of the induced magnetic field with the internal temperature. Different curves stands for different background magnetic field.

Fig 1 only shows the case of paramagnetism, that is no magnetic interaction. Since the problem of magnetism is still an open question in condensed matter physics (Feng & Jin, 2005), we will try to discuss this problem in section V. We will first discuss the upper limit of the magnetic field of magnetars in our theoretical frame. Since this can be tested by observation directly.

## 4. The Upper Limit of Magnetic Field of Magnetars

We note that the temperature factor $f(\mu_n B/kT)$ tends to 1 when the temperature decreases low enough. Actually, $f(\mu_n B/kT) \sim 1$ as long as $\mu_n B/kT \gg 1$. For example, this is true when $T \sim 10^7 K$ if B= $10^{15}$ gauss.

There is an upper limit for the induced magnetic field of the $^3P_2$ superfluid according to eq.(12). It corresponds to the maximum value unity of the temperature factor $f(\mu_n B/kT)$. This upper limit can be realized when all the magnetic moments of the $^3P_2$ neutron Cooper pairs are arranged with the paramagnetic direction as the temperature become low enough. The upper limit of the magnetic field for the magnetars in our point of view is

$$B_{\max}^{(in)}(^3P_2) = \frac{2\mu_n N_A m(^3P_2)}{R_{NS}^3} q \quad gauss$$
$$\approx 2.02 \times 10^{14} \eta \quad gauss \quad (14)$$

Here $\eta = \frac{m(^3P_2)}{0.1 m_{Sun}} R_{NS,6}^{-3} [\frac{\Delta_n(^3P_2)}{0.05 MeV}]^{1/2}$ is the dimensionless factor describes both the macroscopic and microscopic properties of neutron stars.

The maximum magnetic field for magnetars depends on the total mass of the anisotropic neutron superfluid of the neutron star. It is well known that the upper limit of the mass for the neutron stars is more than $2.5 M_{Sun}$. It is therefore possible that the mass of the anisotropic neutron superfluid of the heaviest neutron star may be about. Hence, the maximum of the magnetic field for the heaviest magnetar may be estimated to be $(3.0-4.0) \times 10^{15}$ gauss. This can be tested directly by magnetar observations.

## 5. Conclusions and Discussion

The strong magnetic field of magnetars may originate from the induced magnetic field due to the paramagnetic moment of the $^3P_2$ superfluid with significant mass more than $0.1 M_\odot$ at temperature about $10^7 K$.

The evolutionary scenario of the magnetic field of neutron stars may be depicted as follows:

1) First, the fossil magnetic field of the collapsed core due to the conservation of magnetic flux during supernova explosion would be greatly boosted up more than 90 times to $B^{(0)}$ by the Pauli paramagnetization of the highly degenerate relativistic electron gas just after the formation of the neutron stars (Peng and Tong 2007).

2) The magnetic fields of all neutron stars with a significant $^3P_2$ superfluid (for example, $\eta>1$), would increase gradually when the temperature of the cooling star decreases down to $(T/10^7 K)<2\eta$. Here we provide a choice for a normal neutron star to evolve toward a magnetar.

3) For some magnetars, their magnetic field seem be so strong as $1\times10^{15}$ gauss according to the standard magnetic dipole model. The corresponding magnetic field is calculated by the observed period ($P$) and its varied rate ($dP/dt$) following the standard magnetic dipole model,

$$B_{p,12} \approx 3.3\times 10^7 \sqrt{P\dot{P}} \quad Gauss \quad (15)$$

However, the calculated magnetic field strength may be weaker than the standard value above according to our hybrid model for pulsar spin down (Peng, Huang & Huang, 1982). Taking the contribution of neutrino radiation by the superfluid neutron vortices into account, we obtain

$$P\dot{P} = A_0(B_{p,12}^2 + B_0 G(n)P^3)$$
$$G(n) = \frac{\overline{n^7}}{\overline{n}}, \qquad B_0 = 3.01\times 10^{-8}\sin^{-2}\alpha \quad (16)$$

Where $n$ is the vortex quantum number of both the isotropic and the anisotropic superfluid neutron vortices, which decreases with the increasing period of the pulsar. In our opinion, the observed strong magnetic field strength of the magnetars should be re-estimated and the magnetic field of the magnetars may somewhat be lower than $1\times10^{15}$ gauss.

4) The maximum of the magnetic field for the magnetar is $(3.0-4.0)\times10^{15}$ gauss according our idea.

5) The core temperature of the magnetars is about $10^7$ K in our model. While observations show that some SGR's and AXP's have high thermal-type-spectrum X-ray flux, being among the hottest neutron stars. We shall discuss this question and give reasonable and consistent explanation in our another paper "Fermi energy of electrons in neutron stars with strong magnetic field and magnetars" (Peng et al., 2015, submitted)